\documentclass[12pt,preprint]{aastex}
\begin{document}
\title{Radio halo formation through magneto-turbulent particle
       acceleration in clusters of galaxies}
\author{Hiroshi Ohno \altaffilmark{1}, Motokazu Takizawa \altaffilmark{2,3}, and Shinpei Shibata \altaffilmark{2}}
\altaffiltext{1}{Yamagata Junior College, Yamagata, 990-2316, Japan;
                 ohno@yamagata-jc.ac.jp}
\altaffiltext{2}{Department of Physics, Yamagata University, Yamagata, 990-8560, Japan;
                 takizawa@ksirius.kj.yamagata-u.ac.jp, shibata@ksirius.kj.yamagata-u.ac.jp} 
\altaffiltext{3}{Department of Astronomy, University of Virginia, P.O. Box 3818 Charlottesville, VA 22903-0818}


\begin{abstract}

We developed a magneto-turbulent model for the cosmic ray (CR) electrons
   seen in the radio halo clusters of galaxies.
Steady state momentum distribution functions of the CR
   electrons are calculated for given spectra of the turbulent Alfv\'{e}n waves.
The radio spectrum
    produced by the obtained CR electron distribution
   is compared
   to the observed radio spectrum of the Coma radio halo.
We find that 
    the observed radio spectrum of the Coma cluster is well
    reproduced when the spectral index of the  turbulent
    Alfv\'{e}n waves is $\sim 2.8$.
The obtained energy spectrum of the turbulent Alfv\'{e}n waves
   is steeper than that expected from the turbulence theory,
   suggesting back reaction of the particle acceleration.
The fitting procedure constraints the amplitude of the turbulent Alfv\'{e}n waves
   as well as the spectral index.
Then we estimate the dissipation of the turbulent Alfv\'{e}n waves,
   which is found to be less than the cooling rate by X-ray radiation.
We suggest that
   the turbulence which is sufficient for particle acceleration
   is developed in the clusters containing the radio halo.
It is most likely that cluster mergers create the turbulence
   and seed relativistic electrons.

\end{abstract}

%



%


\section{INTRODUCTION}
The radio halo in a cluster of galaxies is
   a diffuse nonthermal (synchrotron) radio emission
   seen in the central region of $\sim 1$ Mpc in diameter.
According to Hanisch (1982),
   the typical luminosity is
   $ L_{\rm t} \sim 10^{40} - 10^{41}$ erg s$^{-1}$
   (10 MHz $\leq \nu \leq$ 10 GHz),
   and the spectral index is $1 \lesssim \alpha \lesssim 2$.
The typical spatial scale of the radio halos is
   larger than the optical core radius,
   but smaller than the Abell radius.
Giovannini, Tordi, \& Feretti (1999) inspected a sample of
205 clusters from the X-ray-brightest Abell-type clusters
(Ebeling et al. 1996) to search for new radio halo and relic 
candidates. 
They found only 29 candidates. 
In addition, they found that occurrence of the radio halos and relics
is higher in clusters with higher X-ray luminosity and
higher temperature and confirmed that the positive correlation between the
absence of a cooling flow and the presence of radio halo. 
The rarity of the radio halo is one characteristic to be considered
in its formation scenario.

In clusters of galaxies with radio halos, nonthermal
X-ray radiation due to inverse Compton scattering of cosmic 
microwave background (CMB) photons is expected (Rephaeli 1979).
Indeed, nonthermal hard X-ray radiation was recently
detected in a few rich clusters (e.g., Fusco-Femiano et al. 1999;
Rephaeli, Gruber, \& Blanco 1999; Fusco-Femiano et al. 2000)
and several galaxy groups (Fukazawa 2001), although their
origin is still unidentified. 

Among clusters with radio halo, the Coma cluster is the most informative.
The halo size is
   $30'-40'$ in diameter corresponding to $600-800\ h_{100}^{-1}$ kpc.
The radio halo region contains two bright radio galaxies, NGC4874 and NGC4869.
The spatially averaged spectral index of the halo,
   with reduction of contamination by the radio galaxies,
   is $\alpha \sim 1.3$.
Giovannini et al. (1993) presented a spatial distribution
   of the spectral index between 326 MHz and 1.4 GHz.
They found that
   the radio spectrum of the central halo region with size of $\sim 15'$
   is flatter, $\alpha \sim 0.8$,
   while in the vicinity of the tailed radio source associated with NGC4869
   is steeper, $\alpha \sim 1.8$.
This means that
   high energy cosmic rays (CRs) are more abundant
   in the halo region than in the vicinity of the tailed source.
The flatness of the radio halo spectrum suggests
   particle acceleration in the intracluster space.
This has been pointed out by Giovannini et al. (1993).
The radio spectrum of the outer region of the radio halo is
   steeper ($\alpha \sim 1.8$) than that of the central region.

Kim et al. (1990) studied the magnetic fields in the Coma cluster
   by using the rotation measures (RMs) of background radio sources
   (QSO and radio galaxies).
The observed RMs of the background sources seen through the Coma cluster
   have an excess of $\sim$ 38 rad m$^{-2}$.
By adopting a galaxy scale of $10-40$ kpc as a correlation length,
   they found the (electron density weighted) amplitude of the
random magnetic field to be $\sim$ 1 $\mu$G.
On the other hand, 
 the comparison between nonthermal hard X-ray emission and synchrotron radio emission
 gives the volume averaged magnetic field strength,
 if all the hard X-ray is emitted through inverse Compton scattering of CMB photons 
 (Rephaeli 1979).
Rephaeli, Gruber, \& Rothschild (1987) obtained a lower limit value $\sim 0.11~\mu$G
 by using the upper limit on the hard X-ray emission from the Coma cluster.
The first detection of the hard X-ray emission from the Coma cluster by {\it BeppoSAX}
 gives the field strength of 0.15 $\mu$G (Fusco-Femiano et al. 1999).
The obtained values are significantly smaller than those derived from RMs.
However, it is important to note that
 (1) these estimates assume that the emitting volumes of the hard X-ray and radio emission coincide,
    which might not be guaranteed, and that 
 (2) these estimates are affected by the presence of additional infrared background radiation fields
     (Schlickeiser \& Rephaeli 1990).
These simplification can make the derived field strength smaller than 
 that derived from RMs.

The origin of the CR electrons in the clusters with radio halo is
still unclear. 
Jaffe (1977)
   proposed the primary electron model
   where the radio halo is an emission by CR electrons
   diffusing away from radio galaxies in a cluster.
However, this model contains some difficulties,
   one of which is that the diffusion distance does not seem as large as the radio halo extent.
Dennison (1980) proposed the secondary electron model where
the relativistic electrons are produced through decay of
charged pions induced by the interaction between
relativistic protons from radio galaxies and thermal protons 
in ICM. 
In this model, however, too much gamma-ray emission 
is produced at least for the Coma cluster (Blasi \&
Colafrancesco 1999).

Excess of the high energy CR electrons in the halo region,
    indicated by the flat spectrum,
    strongly suggests the particle acceleration in an intracluster space.
Models considering particle acceleration by intracluster magnetic turbulence
   were discussed by Roland (1981), Schlickeiser, Sievers, \& Thiemann (1987), 
   and Petrosian (2001).
In this paper,
   we model the radio halo in terms of
   particle acceleration by the intracluster turbulent magnetic fields.
The turbulence is assumed to be an ensemble of Alfv\'{e}n waves,
   and the CR electrons are accelerated by pitch angle scattering
   by the Alfv\'{e}n waves.
Distribution functions of the CR electrons in energy space
   are obtained by solving a Fokker-Planck equation 
   for the assumed turbulent spectra with various power indices.
The calculated radio spectra are compared with the observed one
   to determine the energy spectrum of the turbulent Alfv\'{e}n waves.

Deiss et al. (1997) proposed a merger shock acceleration model
 which explains the rarity of radio halos.
Numerical simulations by Takizawa \& Naito (2000) show
 radio emission is luminous while merger shocks accelerate CR electrons but decrease rapidly
 because of inverse Compton cooling after the shocks disappear.
The transient feature of this model is consistent with the rarity of radio halos.
We will discuss the relation between our model and theirs.

In \S2,
   the basic equations to calculate a radio spectrum
   are introduced.
In \S3,
   the calculated spectra are interpreted,
   and
   the energy spectra of the turbulent Alfv\'{e}n waves
   is determined
   so as to reproduce the observed radio spectrum.
In \S4,
   the generation mechanism of the obtained energy spectra is investigated,
   and over all turbulence structure is discussed.
\S5 is for summary.

%


\section{MODEL}

\subsection{Resonant Acceleration by Turbulent Alfv\'{e}n Waves}

Energy of CR electrons is gained 
   through a number of pitch angle scattering by Alfv\'{e}n waves
   (e.g., Wentzel 1974; Skilling 1975; Blandford \& Eichler 1987;
     Berezinsky et al. 1990; Terasawa 1991).
The resonance scattering condition 
   for an electron of
    velocity along the magnetic field $v_\parallel$
   with a wave of a frequency $\omega_{\rm A}$
        and wave number $k$
   is
\begin{equation}
 \label{rescon1}
  \omega_{\rm A} =  k v_\parallel + \Omega_{\rm e},
\end{equation}
   where $\Omega_{\rm e}$ is the electron cyclotron frequency,
   and
   the Alfv\'{e}n velocity relates $\omega_{\rm A}$ and $k$ by
   $\omega_{\rm A}=k v_{\rm A}$.
This condition means that
   the electron travels with the Alfv\'{e}n wave
   along the mean field at a locked phase.
The wave number resonant with the CR electron
   of energy $E$ is approximated as
\begin{equation}
   \label{rescon2}
    |k_{\rm res}| 
    =
    \left| \frac{-\Omega_{\rm e}}{ \mu v-v_{\rm A}} \right|,
\end{equation}
   where $\mu$ is the cosine of the pitch angle,
         $v \sim c$ is the CR electron velocity.

When $\mu \gg v_{\rm A}/c$, $| k_{\rm res} | \sim e B_0 / E$,
   where $e$ and $E$ are the charge and the energy of the CR electron,
         and $B_0$ is the large-scale-field strength.
The energy of the CR electron is $E \sim m_{\rm e0} c^2 \gamma$,
   where $m_{\rm e0}$ and $\gamma$ are the rest mass and the Lorentz factor of the CR electron.
For $E \gtrsim 100$ MeV,
   $k_{\rm res} \lesssim 3 \times 10^{-12}B_{\mu {\rm G}}$ cm$^{-1}$,
   where $B_{\mu {\rm G}}$ is the large-scale-field strength
   in units of $\mu$G.
In this wave number range, the magnetohydrodynamical description of the
   Alfv\'{e}n wave is valid because 
   $k \ll k_{\rm max} 
    \sim \Omega_{\rm i} / v_{\rm i}
    \sim 10^{-10}$ $B_{\mu {\rm G}}$ cm$^{-1}$,
 where $v_{\rm i}$ is the ion sound velocity
 (Holman, Ionson, \& Scott 1979).
This picture is not valid when $\mu$ is much less than $v_{\rm A}/c$.
When $\mu \sim 0$, the resonant wave number
   $k_{\rm res} \sim e B_0 / (m_{\rm e} c \gamma v_{\rm A}) \sim 10^{-8}$ cm $^{-1}$,
   which is larger than the $k_{\rm max}$ for 100 MeV electrons,
   and therefore well in the whistler or ion-cyclotron range.
Steinacker \& Miller (1992) discuss the gyroresonance of the electrons
   with parallel transverse cold plasma waves including the Alfv\'{e}n waves,
   whistlers, and cyclotron waves, and derive the Fokker-Planck equation
   describing the evolution of the electron distribution function.
Also, Felice \& Kulsrud (2001) suggest that
   scattering at $\mu \sim 0$ is affected by transit-time damping effect with long-wavelength waves.
These processes should be included in the model for the radio halos.
For this we have to give energy spectra of all wave modes
   related to these processes to obtain diffusion coefficients.
However, it is difficult to assume the energy spectra of all wave modes realistically
   because we do not have sufficient information about 
   the generation mechanisms of the waves and the interaction between the waves.
The aim of this paper is to evaluate the energy spectrum of the Alfv\'{e}n waves
   accelerating the CR electrons producing the radio halo.
Then we attempt to do this 
   assuming that pitch angle scattering by Alfv\'{e}n waves is dominant process,
   hoping corrections by other processes near $\mu \sim 0$ is small.
However, we must note that
   if the correction to the present form of diffusion coefficient by other wave modes is significant,
   our evaluation of the energy spectrum of the Alfv\'{e}n wave mode would not be valid.
In this paper,
   we adopt a pitch-angle averaged diffusion coefficient of the resonant scattering
   as a reasonably good approximation (Isenberg 1987; Schlickeiser 1989; Blasi 2000).

The energy spectrum of the turbulent Alfv\'{e}n waves is assumed to be
   a single power law with an index $w >1$:
\begin{eqnarray}
 \label{magene}
  P(k) = b \frac{B_0^2}{8 \pi} \frac{(w-1)}{k_0} 
          \left ( \frac{k}{k_0} \right ) ^{-w},
  & k_0 \leq k \leq k_{\rm max},
\end{eqnarray}
   where the applicable range,
    $k_0 \leq k \leq k_{\rm max}$,
   is discussed later;
   the normalization factor $b$ indicates
   the fractional energy density of the turbulent Alfv\'{e}n waves
   if $k_{\rm max} \gg k_0$.
In this case, the pitch angle-averaged momentum diffusion coefficient is given by
   \begin{equation}
    \label{diffpp}
     D_{\rm pp} =  (m_{\rm e0} c)^2 a_1 \gamma^w,
   \end{equation}
where
   \begin{equation}
    \label{a1}
      a_1 = \frac{w-1}{w(w+2)} \pi b 
            \frac{v_{\rm A}}{c} k_0 v_{\rm A}
            \gamma_0^{2-w},
   \end{equation}
   (Isenberg 1987; Schlickeiser 1989; Blasi 2000).
Here the relation between $\gamma$ and $k$ is
   \begin{equation}
    \label{k-g}
      k = \frac{e B_0}{m_{\rm e0} c^2 \gamma},
   \end {equation}
   and $\gamma_0$ corresponds to $k_0$.
Also the spatial and momentum diffusion coefficients are related by
   \begin{equation}
    \label{diff}
     D_0 = \frac{v_{\rm A}^2}{9} \frac{p^2}{D_{\rm pp}},
   \end{equation}
   (Melrose 1980; Schlickeiser et al. 1987).

The transit-time damping effect of fast magnetosonic waves
   plays a more crucial role for lower energy ($\gamma < 1836$) electrons
   (Schlickeiser \& Miller 1998; Ragot \& Schlickeiser 1998),
   which will emit EUV radiation through inverse Compton scattering of CMB photons
   and possibly contribute in lower energy ends of the radio and hard X-ray emission.
Therefore, this effect should be considered when broader energy band spectrum models are constructed.

\subsection{Basic Equations}

We simplify the problem as follows.
(1) The distribution of the CR electrons is isotropic
    and depends only on the absolute value of the momentum $\mid$ {\boldmath$p$} $\mid$.
(2) Spatial averaged distribution is concerned.
(3) Steadiness is assumed for simplicity;
    reality of this assumption is discussed later.
Thus, the distribution function has a form $f=f(p)$.
The effect of the spatial diffusion is represented by
   the escape time formalism in the form of $-f(p)/T_{\rm d}$,
   where the diffusion time is estimated by
\begin{equation}
 \label{difft}
  T_{\rm d}(p) = \frac{R^2}{D_{\rm 0}},
\end{equation}
   where $R$ is the spatial scale,
   i.e., the size of a radio halo.

The Fokker-Planck equation for the present model is, therefore,
\begin{eqnarray}
 \label{basic1}
        \frac{1}{p^2} \frac{d}{d p}
        \left \{ D_{\rm pp} p^2 \frac{d f}{d p}
        - \dot{p}_{\rm loss} p^2 f \right \} & \nonumber \\ 
    - \frac{f}{T_{\rm d}} + Q(p) & = 0,
\end{eqnarray}
   where $\dot{p}_{\rm loss}$ indicates the momentum loss
   due to synchrotron radiation and inverse Compton effect,
   and $Q(p)$ is the particle injection rate at low energies.
For the loss term, we use
\begin{equation}
 \label{ploss}
  \dot{p}_{\rm loss} = -\frac{4 \sigma_{\rm T}}{3 m_{\rm e0}^2 c^2}
                            [ U_{\rm mag} + U_{\rm ph} ] p^2,
\end{equation}
   where $U_{\rm mag} = B_0^2/(8 \pi)$ is the energy density of
   the large-scale fields,
   and
   $U_{\rm ph} \sim 4 \times 10^{-13}$ erg cm$^{-3}$ is
   the energy density of 2.7 K background photons.
For $Q(p)$,
   the mono-energetic injection at $\gamma = \gamma_{\rm i}$ is assumed:
\begin{equation}
 \label{source}
  Q[p(\gamma)] = q_{\rm i} \delta [p(\gamma)-p(\gamma_{\rm i})],
\end{equation}
   where $q_{\rm i}$ is
    the injection rate averaged over the radio halo region.

We rewrite equation (\ref{basic1}) in terms of $\gamma \sim p/(m_{\rm e0} c$),
   in stead of $p$, and extract factors which are independent of $\gamma$
   to get
\begin{eqnarray}
 \label{basic2}
   a_1 \gamma^w \frac{d^2 f}{d \gamma^2}
      +(w+2) a_1 \gamma^{w-1} \frac{d f}{d \gamma} & \nonumber \\
   + \frac{1}{\gamma^2} \frac{d}{d \gamma}
      \{ a_2 \gamma^4 f \}-\frac{f[p(\gamma)]}{T_{\rm d}} + Q[p(\gamma)]
   & =0
\end{eqnarray}
   where,
\begin{equation}
 \label{a2}
  a_2 = \frac{4 \sigma_{\rm T}}{3 m_{\rm e0} c}
                            [ U_{\rm mag} + U_{\rm ph} ].
\end{equation}
The first term of equation (\ref{basic2}) represents
   simple diffusion in the $\gamma$-space,
   while the second term causes the upward advection in the
$\gamma$-space, i.e., the acceleration,
   with `velocity' $\dot{\gamma}_{\rm accel}= (w+2) a_1 \gamma^{w-1}$.

\subsection{Characteristic Timescales}

Two other characteristic timescales than $T_{\rm d}$ are used to normalize the equation:
   one is the acceleration time
\begin{equation}
 \label{ta}
  T_{\rm a}\sim \frac{\gamma}{\dot{\gamma}_{\rm accel}}
               =    \frac{\gamma^{2-w}}{(w+2)a_1},
\end{equation}
   and the other is the momentum loss time
\begin{equation}
 \label{tl}
  T_{\rm l} \sim 
   \frac{p}{|\dot{p}_{\rm loss}|} = \frac{1}{a_2 \gamma}.
\end{equation}
With the help of equations (\ref{diffpp}), (\ref{a1}) and (\ref{diff}),
   the spatial diffusion time (eq. [\ref{difft}]) is rewritten as
\begin{equation}
 \label{td}
  T_{\rm d} = \frac{9 R^2}{v_{\rm A}^2} a_1 \gamma^{w-2}.
\end{equation}
From equations (\ref{ta}) and (\ref{td}),
   $T_{\rm d} \propto 1/T_{\rm a}$.
This is because the larger the number of scattering is,
   the higher the acceleration rate is
   and
   the shorter the diffusion distance is.
Thus,
   the shorter the acceleration time is,
   the larger diffusion time is.
By using these three timescales,
   equation (\ref{basic2}) can be rewritten as follows,
\begin{eqnarray}
  \label{basic3}
  \left( \frac{T_{\rm l}}{T_{\rm a}} \right) \left ( \frac{\gamma^2}{w+2} \right ) \frac{d^2 f}{d \gamma^2}
  +\left( \frac{T_{\rm l}}{T_{\rm a}} \right) \gamma \frac{d f}{d \gamma}   & \nonumber \\
  +\gamma^{-3} \frac{d}{d \gamma} (\gamma^4 f) 
  -\left( \frac{T_{\rm l}}{T_{\rm d}} \right) f + T_{\rm l} Q  & =0.
\end{eqnarray}

To understand the behavior of the solution,
   it is convenient to use a critical energy $m_{\rm e0} c^2 \gamma_{\rm c}$ 
   with which $T_{\rm a}=T_{\rm l}$. 
The Lorentz factor for the critical energy is
\begin{equation}
 \label{gc}
  \gamma_{\rm c} = \left[ \frac{(w+2)a_1}{a_2} \right]^{\frac{1}{3-w}}.
\end{equation}
By using $\gamma_{\rm c}$,
   $T_{\rm l}/T_{\rm a}$ and $T_{\rm l} / T_{\rm d}$ can be written as follows
\begin{equation}
 \label{tlta}
  \frac{T_{\rm l}}{T_{\rm a}} = \left ( \frac{\gamma}{\gamma_{\rm c}} \right )^{w-3},
\end{equation}
\begin{equation}
 \label{tltd}
  \frac{T_{\rm l}}{T_{\rm d}} = \frac{(w+2)}{9} \left ( \frac{v_{\rm A}}{R a_2} \right )^2
                                \gamma_{\rm c}^{-2} \left ( \frac{\gamma}{\gamma_{\rm c}} \right )^{1-w}.
\end{equation}
The large-scale-field strength and the spatial scale are given for each cluster,
   and hence $a_2$ is regarded as a constant.
Therefore, the model free parameters are
   $\gamma_{\rm c}$, $w$, and $q_{\rm i}$.

When $w=2$, the acceleration time (\ref{ta}) and the diffusion time (\ref{td}) are
   independent of $\gamma$.
In this case, the Fokker-Planck equation has an analytical solution (Schlickeiser 1984).
Schlickeiser et al. (1987) applied an approximated form of this analytical solution
    to the radio spectrum of the Coma halo.
For solar flares,
   analytical solutions of the Fokker-Planck equation in general cases with $w \neq 2$ and finite $T_{\rm d}$
   are discussed in Dr\"{o}ge \& Schlickeiser (1986) and Steinacker, Jaekel, \& Schlickeiser (1993).
Park \& Petrosian (1995) discussed the effects of the boundary conditions
   on the analytical solutions of the Fokker-Planck equation in general cases.
However, these models cannot be applied to our model in their original forms
   because the synchrotron and the inverse Compton losses are neglected there.

One of our main purposes in this paper is to search the wider range of parameter space
   than the former work to investigate the radio spectra under various conditions.
For this purpose, it is convenient to use numerical methods to solve the Fokker-Planck equation
   in general cases with $w \neq 2$ and the radiation loss.
Equation (\ref{basic2}) is one dimensional diffusion equation
   and is reduced to the algebraic equation of $f$
   by the differencing method.
The obtained equation is numerically solved by the Gaussian elimination.
The boundary conditions are
   $d f/d p=0$ at $p \rightarrow 0$
   and
   $f \rightarrow 0$ at $p \rightarrow \infty$.
The comparison between the obtained numerical solution and the analytic one for $w=2$
 suggests that the errors are $\sim$ 1 \% in the present calculation.

\subsection{Radio Flux}

We calculate the radio spectrum from the obtained electron distribution function.
The volume emissivity is given by
\begin{equation}
 \label{epsnu}
  \epsilon_{\nu} = \int P_{\rm s}(p, \nu) f(p) p^2 \mbox{d} p,
\end{equation}
   where $P_{\rm s}$ is the synchrotron radiation spectrum due to a single electron
   of momentum $p$:  
\begin{equation}
 \label{ps}
   P_{\rm s}(p, \nu) = \frac{\sqrt{3} e^3}{m_{\rm e0} c^2}B_0 x
                       \int^{\infty}_x K_{\frac{5}{3}}(\eta) \mbox{d} \eta,
\end{equation}
   with
   $x = \nu / [3 e B_0 \gamma^2 / (4 \pi m_{\rm e0} c)]$.
The peak of this spectral distribution occurs at a frequency
\begin{equation}
 \label{freq}
   \nu = \nu_{\rm B} \gamma^2,
\end{equation}
   where $\nu_{\rm B}= (0.29 \times 3  e B_0)/(4 \pi m_{\rm e0} c)$.
This relation (eq. [\ref{freq}]) defines
   a critical frequency $\nu_{\rm c}$ corresponding to 
   the critical energy $m_{\rm e0} c^2 \gamma_{\rm c}$.

The radio flux is
\begin{equation}
 \label{snu}
  S_{\nu} = \epsilon_{\nu} l \Omega_{\rm source},
\end{equation}
   where $l$ is the source thickness,
         $\Omega_{\rm source}$ is the solid angle of the radio halo.
The radio flux is calculated numerically from the obtained distribution function
   to compare the observation.
In \S 3,
   the radio flux is calculated for the radio halo in the Coma cluster,
   where we assume
   $ l (=2R) = 600$ kpc,
   $ \Omega_{\rm source} = 30' \times 30'$
   by adopting the size of the radio halo
   $R = 300$ kpc.

\subsection{Model Parameters}
The size of the radio halo $R=300$ kpc,
   the gas number density $n=10^{-3}$ cm$^{-3}$
   are fixed for the Coma cluster,
   and
   we examine the case with two field strength $B_0 =$ 0.1 and 1 $\mu$G.
The model free parameters are $\nu_{\rm c}$, $w$, and $q_{\rm i}$.

The injection energy
   is assumed to be in the range 100 MeV $\lesssim E_{\rm i}
\lesssim$ 1 GeV.
In typical intracluster conditions, the cooling time of
   $\sim 100$ MeV electrons is fairly long ($\sim$ a few Gyr). 
Thus, it is quite natural that there are abundant
   electrons with this energy range produced by some past
   active phenomena (Sarazin 1999). 
The source electrons may be produced at shocks in ICM
   (Takizawa \& Naito 2000) or leaked out from 
   radio galaxies, or normal galaxies. 
We choose the value of $q_{\rm i}$ so that
   the calculated net flux fits to the observed one.


\section{RADIO SPECTRUM}
\subsection{Radio Spectrum and Alfv\'{e}n Wave Energy Spectrum}
\begin{figure}
\epsscale{0.8}
\plotone{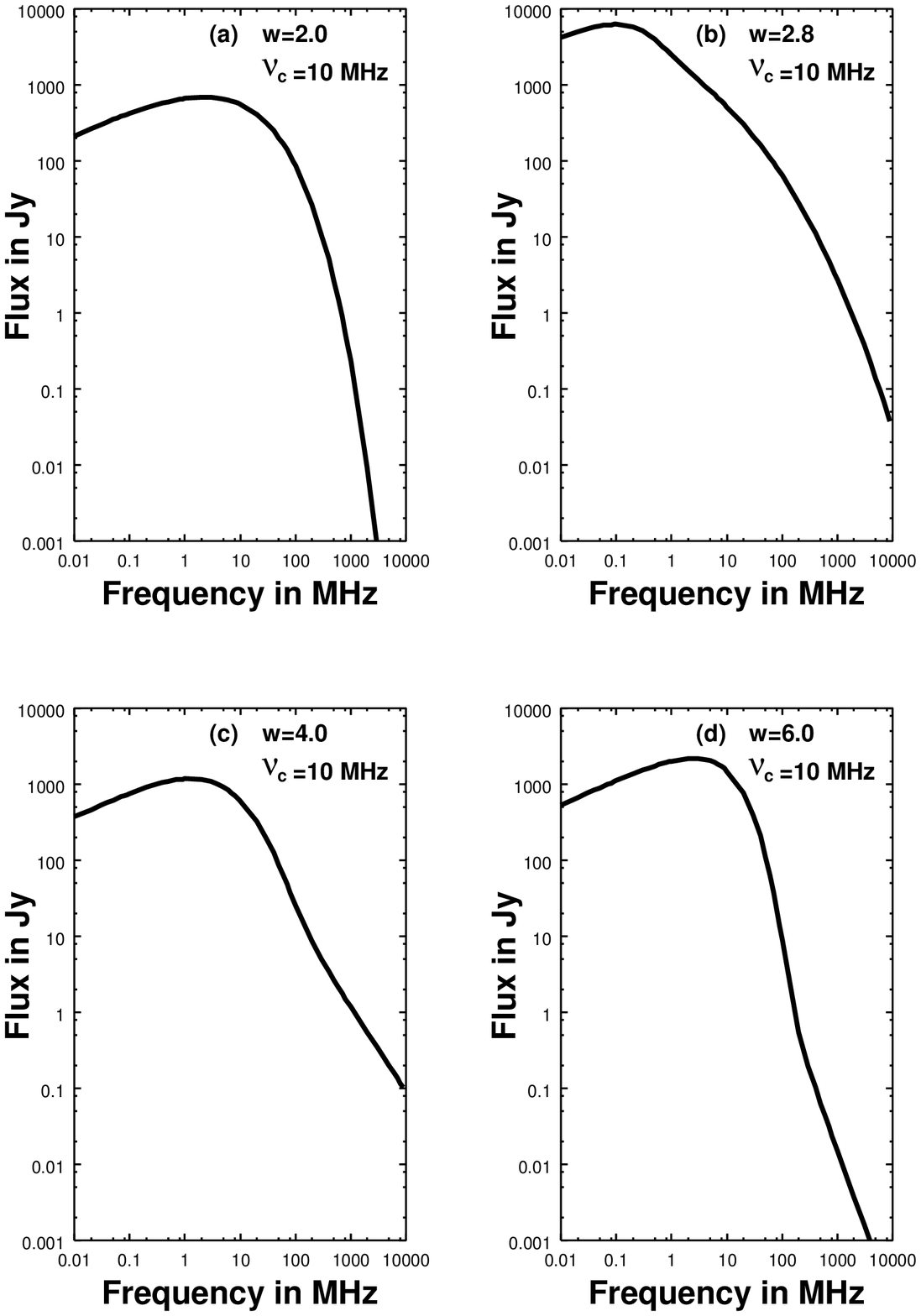}
\caption
{Calculated radio spectrum with $B_0=1~\mu$G, $R=300$ kpc, $\nu_{\rm c}=10$ MHz
 and various values of $w$. (a)$w=2$; (b)$w=2.8$; (c)$w=4$; (d)$w=6$.
 The radio spectrum shows convex and concave shape for $w<3$ and $w>3$, respectively.
 \label{fig1}}
\end{figure}

Before comparison with the observations,
   we shall see parameter dependence of the solution
   to clarify the physical properties of the model.
Four examples of the calculated radio spectrum are taken up in Figure \ref{fig1}
   with different indices (a)$w=2$, (b)$w=2.8$, (c)$w=4$, (d)$w=6$
   and with the fixed critical frequency at $\nu_{\rm c}=10$ MHz.
The vertical axis of the figure is the radio flux in units of Jy,
   and 
   the horizontal axis is the frequency in MHz.
We assume $B_0=1$ $\mu$G,
   and
   $q_{\rm i}$ is chosen 
   so that the total luminosity in the 10 MHz$-$10 GHz range becomes
   $L \sim 10^{41}$ erg s$^{-1}$ for each example.
The injection energy is $E_{\rm i} \sim 100$ MeV 
   which corresponds to the frequency $\nu \sim 0.05$ MHz
   in the models of $w=$2 and 2.8,
   and $E_{\rm i} \sim 1$ GeV 
   corresponding to $\nu \sim 5$ MHz
   in the models of $w=$4 and 6.

The shape of the distribution function is determined through competition between acceleration and loss.
So, the useful parameter is $T_{\rm l} / T_{\rm a}$ (see eq.[\ref{tlta}]).
The ratio $T_{\rm l} / T_{\rm a}$ is proportional to $\gamma^{w-3}$,
   and therefore characteristic features of the obtained distribution function
   and, in turn, the shape of the radio spectrum
   depend on whether $w$ is larger or smaller than 3.
When $w<3$,
   the injected electrons are scattered up to
   the critical energy $\gamma_{\rm c} m_{\rm e0} c^2$ 
   because $T_{\rm a} < T_{\rm l}$ for $\gamma<\gamma_{\rm c}$,
   whereas electrons diffusing out to
   energies higher than $\gamma_{\rm c} m_{\rm e0} c^2$
   flow back in the energy space owing to the loss ($T_{\rm a}>T_{\rm l}$).
Therefore,
   the electrons tend to gather around the critical energy $\gamma_{\rm c} m_{\rm e0} c^2$.
For $w=2$, we see in Figure \ref{fig1}a that
   the radio spectrum has the peak at $\nu_{\rm c}$ 
   because of this effect.
For $w=2.8$,
   the effect is attenuated by momentum space diffusion,
   but we can still see `the knee' at $\nu_{\rm c}$.

When $w>3$,
   electrons with energies smaller than $\gamma_{\rm c} m_{\rm e0} c^2$
   essentially flow toward lower energies
   because $T_{\rm a} > T_{\rm l}$ for $\gamma < \gamma_{\rm c}$.
However, a small portion of the electrons is scattered out to 
   energy larger than $\gamma_{\rm c} m_{\rm e0} c^2$.
Once the energy reaches $\gamma_{\rm c} m_{\rm e0} c^2$,
   the electrons are accelerated because  $T_{\rm a} < T_{\rm l}$.
Moreover, the higher electron energy is, 
   the more efficiently electrons are accelerated.
As a result,
   the spectrum shows a concave shape
   around $\sim \nu_{\rm c}$ as seen in Figures \ref{fig1}c and \ref{fig1}d.

We find that
   for $w>3$ the radio spectrum becomes steeper as $w$ is larger.
Figures \ref{fig1}c and \ref{fig1}d show that
   the spectral index for $\nu > \nu_{\rm c}$ is $\sim 1$  and $\sim 2$
   when $w=4$ and $w=6$, respectively.
This is due to strong energy dependence of the acceleration time.
The net acceleration velocity in momentum space is $\dot{p} \sim p/T_{\rm a}$
   because $T_{\rm a}$ is shorter than $T_{\rm l}$ for $\nu > \nu_{\rm c}$.
For a steady state solution,
   the flow flux in momentum space is nearly constant,
\begin{equation}
 \label{fflux}
  4 \pi p^2 \dot{p} f(p) = 4 \pi p^3 \frac{f(p)}{T_{\rm a}} \sim {\rm constant}.
\end{equation}
Hence,
   from equations (\ref{ta}) and (\ref{fflux}), we obtain
\begin{equation}
   f(p) \propto  p^{-(w+1)}.
\end{equation}
By using the monochromatic approximation,
   when the electron distribution function has power law index $\Gamma = w+1$,
   the radio spectrum follows
\begin{equation}
   S_{\nu} \propto \nu^{-(\Gamma-3)/2} = \nu^{-(w-2)/2}.
\end{equation}
This relation explains the radio spectrum becomes steeper as $w$ is larger,
   and $S_{\nu} \propto \nu^{-1}$ and $\propto \nu^{-2}$ when $w=4$ and 6, respectively.

When $w=3$,
   $T_{\rm a}$ and $T_{\rm l}$ have the same $\gamma$-dependence,
   and therefore
   the ratio $T_{\rm l}/T_{\rm a}$ is constant at any $\gamma$
   so that
   there is no critical frequency $\nu_{\rm c}$
   (i.e., $\gamma_{\rm c}$):
   we find that
   $f(p)$ and the radio spectrum follows a single power law.
The larger the value $T_{\rm l}/T_{\rm a}$,
   the flatter the spectral index is.
When $T_{\rm l}/T_{\rm a} \sim 1$,
   the spectral index is $\alpha \sim 1$ in the 30 MHz$-$5 GHz range.
When $w=3$, the static solution to the Fokker-Planck equation
   without the diffusion loss term and the source term is
   $f \propto p^{-a_2/a_1} = p^{-5 T_{\rm a}/T_{\rm l}}$ (Eilek \& Henriksen 1984).
For $T_{\rm a}/T_{\rm l} \sim 1$, $f \propto p^{-5}$, and in turn, $S_{\nu} \propto \nu^{-1}$
   by using the monochromatic approximation.

If $w \neq 3$,
   the critical frequency $\nu_{\rm c}$
   is observable in a radio spectrum
   and
   characterize a radio spectrum.
Determination of $\nu_{\rm c}$ and the spectral index allows us to evaluate
   the energy spectrum of the turbulent Alfv\'{e}n waves
   as shown in the next subsection and \S4.

Table \ref{tab1} shows the result of the parameter survey in the $\nu_{\rm c} - w$ space,
   where $w$ is in between 2 and 6, $\nu_{\rm c}$ between 0.01 MHz and 10 GHz, 
         $B_0 = 1~\mu$ G, and $E_{\rm i}=100$ MeV.
When $\alpha$ falls in the observed range $1 \lesssim \alpha \lesssim 2$ in between 30 MHz and 5 GHz,
   the values are given in the columns.
The model spectra which is too steep with $\alpha>2$ or which is too flat with $\alpha<1$
    are represented by `S' or `F', respectively.
When the spectrum has a peak in the observed frequency range,
  the model is denoted by `P'.
When $w\geq4$, the lower bound of the critical frequency is not known.

We also calculate for the models with $B_0=0.1$ $\mu$G.
The relation between $\alpha$ and the model parameters
   are not altered significantly.

\subsection{The Energy Density of Turbulent Alfv\'{e}n Waves}
\begin{figure}
\plotone{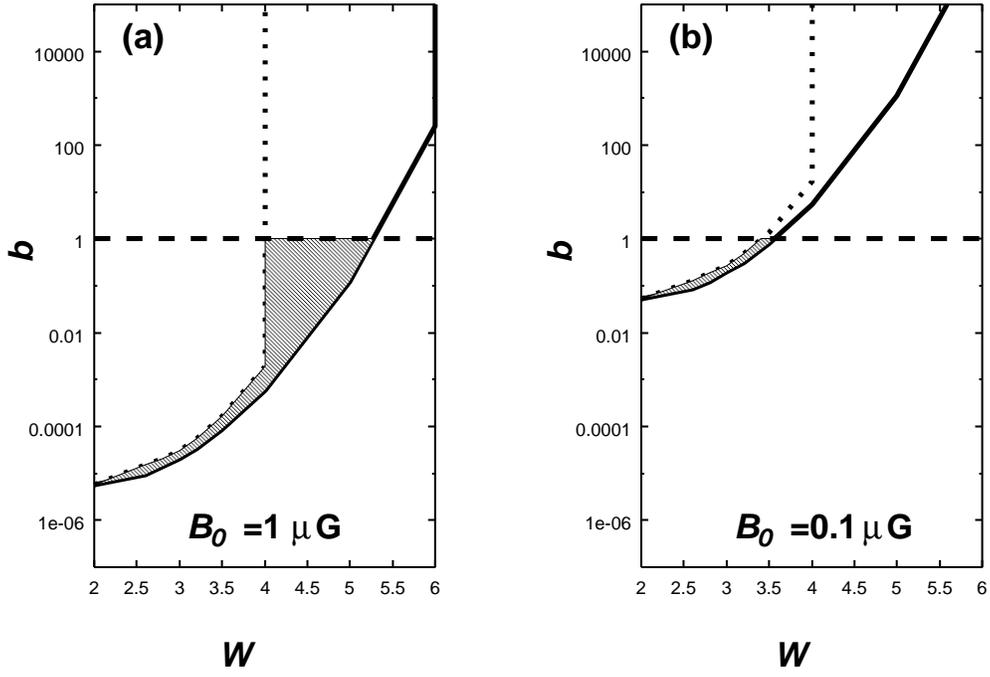}
\caption
{Relation between $b$ and $w$.
 The solid line and the short-dashed line
   indicates, respectively,
   the lower and upper bound of $b$,
   which corresponds to the range $1 \lesssim \alpha \lesssim 2$.     
 The long-dashed line represents the upper limit $b=1$.
 The shaded region indicates the allowed range of $b$ and $w$.
 In the models with $b > 1$,  the energy density of the turbulent Alfv\'{e}n waves exceeds $B_0^2 / (8 \pi)$.
 \label{fig2}}
\end{figure}

The value $b B_0^2 /(8 \pi)$ means the energy density of the turbulent Alfv\'{e}n waves for acceleration 
   of the CR electrons radiating at $\nu \leq$ 5 GHz,
   when we assume $k_0 = k_{\rm 5 GHz}$ in equation (\ref{magene}).
The wave number $k_{\rm 5GHz} \sim 9 \times 10^{-15}$ $B_{\mu {\rm G}}^{3/2}$ cm$^{-1}$ is defined as
   the wave number resonating to the CR electrons radiating at $\nu =$ 5 GHz,
   which is similar to the observed maximum frequency.
Figures \ref{fig2}a and \ref{fig2}b show the relation between $b$ and $w$
   with $B_0=$ 1 $\mu$G and 0.1 $\mu$G, respectively.
The solid line and the short-dashed line
   indicates, respectively,
   the lower and upper bound of $b$,
   which corresponds to the range $1 \lesssim \alpha \lesssim 2$.     

The upper limit of the energy density of the turbulent Alfv\'{e}n waves
   is the energy density of the large-scale-field $B_0^2 / (8 \pi)$.
In figures \ref{fig2}a and \ref{fig2}b,
   the long-dashed line represents the upper limit $b=1$.
The shaded region indicates the allowed range of $b$ and $w$.
When $w < 4$ with $B_{\rm 0}=1~\mu$G,
   $b < 1$, i.e., the condition is met.
However, when $w>4$, the range of the parameters is constrained by the upper limit.
When $B_{\rm 0} = 0.1~\mu$G, the condition is met in the models with $w \lesssim 3.5$.
In the models with $b > 1$,  the energy density of the turbulent Alfv\'{e}n waves exceeds $B_0^2 / (8 \pi)$.
Such models are unrealistic.

As noted in the previous subsection, when $w > 4$,
   the lower bound of the critical frequency is not known.
The values of the lower bound of the critical frequency determined by the constraint $b=1$ are 
   $3 \times 10^{-6}$ MHz and 0.1 MHz for $w=4$ and 5, respectively,
   when $B_0=1~\mu$G.
When $\nu_{\rm c}$ is less than the lower bound,
   the energy density of the turbulent Alfv\'{e}n waves exceeds $B_0^2/(8\pi)$.

\subsection{The Effect of Spatial Diffusion}

The ratio $T_{\rm l}/T_{\rm d}$ (eq.[\ref{tltd}]) is rewritten as follows,
\begin{equation}
\frac{T_{\rm l}}{T_{\rm d}} \sim 3 \times 10^{-2} (w+2) \left( \frac{\gamma}{10^3} \right)^{1-w},
\end{equation}
   where we use $B_0 = 1~\mu$G, $R = 300$ kpc, the gas number density $n = 10^{-3}$ cm$^{-3}$, and 
   $\gamma_{\rm c}=10^3$.
Also, the ratio $T_{\rm a}/T_{\rm d}$ is written as
\begin{equation}
\frac{T_{\rm a}}{T_{\rm d}} = \left( \frac{T_{\rm a}}{T_{\rm l}} \right) \left( \frac{T_{\rm l}}{T_{\rm d}} \right)
                            = \left(\frac{\gamma}{\gamma_{\rm c}} \right)^{3-w} \frac{T_{\rm l}}{T_{\rm d}}
                            \sim 3 \times 10^{-2} (w+2) \left( \frac{\gamma}{10^3} \right)^{4-2w}.
\end{equation}
The spatial diffusion term does not play a significant role in most cases
   because $T_{\rm d}$ is much longer than $T_{\rm a}$ and $T_{\rm l}$
   within the observed frequency range ($\gamma > 10^3$).
However, this is not the case when $w \lesssim 2$, 
   which may be included in our parameter survey,
   e.g., $w = 5/3$ for Kolmogorov.
If, for instance, $w=2$, 
   then $T_{\rm l}/T_{\rm d} \sim 0.1 (\gamma/10^3)^{-1}$ and
   $(T_{\rm a}/T_{\rm d}) \sim 0.1$.
Therefore,
   $T_{\rm d}$ is comparable to $T_{\rm l}$ at a energy range in which we are concern,
   and to $T_{\rm a}$ at all energies.

In \S3.4 we search for the best fit model for the Coma radio halo
   and obtain the best model at $w=2.8$, $\nu_{\rm c}=10$ MHz with $B_0=1~\mu$G.
The spatial diffusion time of this best model is
   $T_{\rm d} \sim 10^{11}$ yr at $\gamma = 6.4 \times 10^4$ corresponding to 5 GHz.
Since the spatial diffusion time is longer than the Hubble time,
   the distribution function does not reach to the steady state.
However, the acceleration time and the energy loss time are shorter than the Hubble time,
   and therefore the distribution function becomes similar to the steady state solution quickly
   but only its normalization is different.
The obtained physical quantities other than the injection rate are fairly valid.
The injection rate $q_{\rm i}$ would be underestimated.

\subsection{Comparison to the Observed Spectrum}

\begin{figure}
\plotone{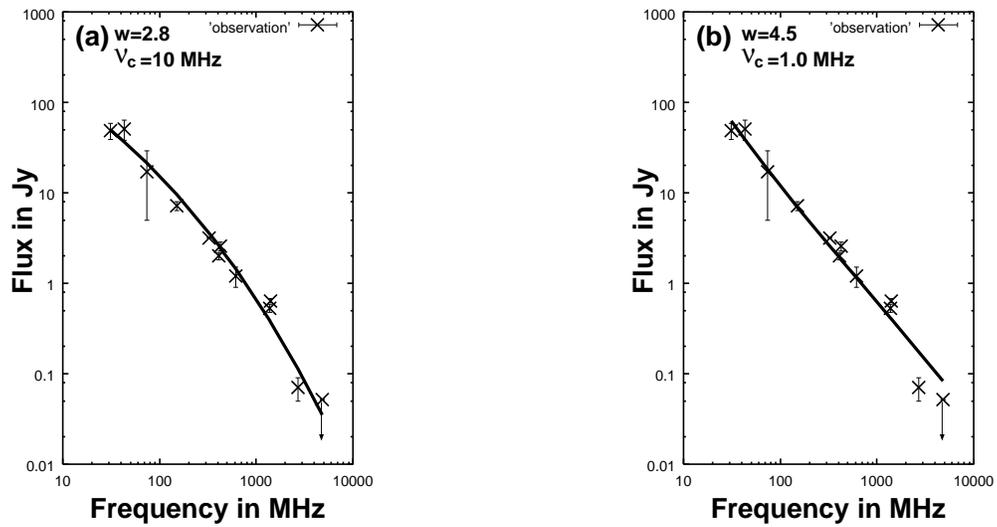}
\caption
{Two examples reproducing the observed radio spectrum.
     (a) $B_0=1~\mu$G, $\nu_{\rm c}=10$ MHz, $w=2.8$;
     (b) $B_0=1~\mu$G, $\nu_{\rm c}=1$ MHz, $w=4.5$;
 The observed intensity is drawn from Deiss et al. (1997).
 \label{fig3}}
\end{figure}

We search for the best fit model for the Coma radio halo
   by the method of least squares.
The average value of squared residuals is calculated as
   \begin{equation}
    \label{j}
     J = \frac{1}{N_{\rm obs}}\sum ( \log S_{\rm cal} - \log S_{\rm obs} )^2,
   \end{equation}
   where $S_{\rm cal}$ is the calculated radio flux
   and $S_{\rm obs}$ is the observed one drawn from Deiss et al. (1997)
   and the number of the data points $N_{\rm obs}$ is 12.
The radio flux (eq.[\ref{snu}]) is calculated with
   the solid angle $ \Omega_{\rm source} = 30' \times 30'$ and the source thickness $ l~(=2R) = 600$ kpc.
(The distance to the Coma cluster is assumed to be 70 Mpc
   by adopting the Hubble constant $H_0 = 100$ km s$^{-1}$ Mpc$^{-1}$.)
The obtained values of $J$ with $B_0=1~\mu$G are presented in Table \ref{tab2}.
The best fit model ($J=0.018$) is obtained when $\nu_{\rm c}=10$ MHz and $w$=2.8.
Figure \ref{fig3}a shows the radio spectrum of this model with $B_0=1~\mu$G,
   and some physical quantities are summarized in Table \ref{tab3}.
For the best fit model,
   $b \sim 1.7 \times 10^{-5}$
   and the energy density of the turbulent Alfv\'{e}n waves for the acceleration is
   $\sim 7 \times 10^{-19}$ erg cm$^{-3}$.
(The factor $b$ represents the fractional energy density
   of Alfv\'{e}n waves accelerating the CR electrons
   radiating at $\nu \leq 5$ GHz [cf. \S 3.2].)
This required energy density
   is much less than
   the energy density of the large-scale-field in the cluster
   $B_0^2 / (8 \pi) \sim 10^{-14}$ erg cm$^{-3}$.

Also, we obtain small values of $J$ when $w=4.5$ and $\nu_{\rm c} \le 1$ MHz.
The radio spectrum with $w=4.5$ and $\nu_{\rm c}=1.0$ MHz is shown in Figure \ref{fig3}b.
The required injection rate $q~(=4 \pi p_{\rm i}^2 q_{\rm i}) \sim 3 \times 10^{-9}$ cm$^{-3}$s$^{-1}$ becomes very large
   when $E_{\rm i}=100$ MeV.
An injection at higher energy might resolve this difficulty.

Figure \ref{fig3}
   shows that
   the spectra with a small value of $J \sim 0.02$,
   can have different shapes.
If the spectrum shows exponential cut off
   such as shown in Figure \ref{fig3}a,
   the index $w$ is smaller than 3,
   or otherwise,
   if the radio spectrum has a concave shape,
   the index $w$ is larger than 3.
When $w=3$, the radio spectrum follows a single power law as mentioned in \S 3.1.
Deiss et al. (1997) suggests that
   the value of the spatially integrated flux above 1.4 GHz is underestimated
   because of the small integration area.
If more diffuse radio power is observed above 1.4 GHz,
   the model with $w>3$ may well explain the observation.
Further observation to determine the radio spectrum at higher frequencies
   set a stronger constraint on the energy spectrum of the turbulent Alfv\'{e}n waves.

When $B_0=0.1~\mu$G,
   the models with ($w,~\nu_{\rm c}$[MHz])=(2.8, 10), and (4.5, 1)
   reproduce the observed spectrum.
Table \ref{tab3} shows that
   the required fractional magnetic energy density $b$
   is larger than that in the case of $B_0=1~\mu$G.
When $w=4.5$ and $\nu_{\rm c}=1.0$ MHz,
   the energy density of the Alfv\'{e}n waves for acceleration exceeds $B_0^2/(8\pi)$
   as noted in \S 3.2.


\section{DISCUSSION}
\subsection{The Energy Spectrum of Turbulent Alfv\'{e}n Waves}
Let us consider physical status of the turbulent Alfv\'{e}n waves in
   our best fit model. Ruzmaikin \& Shukurov (1982) gives the rate of energy-transfer 
   via the nonlinear interaction among different wave numbers as
\begin{eqnarray}
 \left ( \frac{dE}{dt} \right )_{\rm nl}
  & \sim &
   \left. \frac{P(k)^2 k^3}{\rho v_{\rm A}}, \right.\label{wavenl} 
\end{eqnarray}
   where $\rho$ is the mean gas density.
If the energy spectrum of the turbulent Alfv\'{e}n waves is determined
   only through this energy cascade process, it becomes
   $P(k) \propto k^{-3/2}$ in a steady state because $(dE/dt)_{\rm nl}$ 
   is constant for $k$ (Ruzmaikin \& Shukurov 1982).
However, this is significantly flatter than that in our best fit model whose exponent is $-2.8$. 
Thus, other physical processes might work to steepen the Alfv\'{e}n wave spectrum. 
Obviously, the Alfv\'{e}n waves lose their energy through the particle acceleration.
The CR electrons between $p$ and $p+dp$ absorb the energy of Alfv\'{e}n waves at the rate,
\begin{eqnarray}
 \left ( \frac{dE}{dt} \right )_{\rm accel}
  &   =  &
     4 \pi p^2 (m_{\rm e0} c^2) \dot{\gamma}_{\rm accel} f(p) dp  \nonumber \\
  & \sim &
     4 \pi (m_{\rm e0} c^2) p^3 \frac{\gamma}{T_{\rm a}}f(p) \nonumber \\
  & \sim &
     4 \pi m_{\rm e0}^4 c^5 (w+2) a_1 \gamma^{w+2} f(p), \label{waveacc}
\end{eqnarray}
   where we use $dp \sim p$, $\dot{\gamma}_{\rm accel}=\gamma/T_{\rm a}$.
We checked whether this plays a significant role or not in our best fit model. 
For simplicity, we approximate $P(k) \propto k^{-3}$ and $f(p) \propto p^{-6}$. 
In this case, $(dE/dt)_{\rm nl} \propto k^{-3}$ and $(dE/dt)_{\rm accel} \propto k^{-1}$ 
   because the resonant wave number is proportional to $p^{-1}$. 
Therefore, the energy loss due to the particle acceleration is more significant in higher wave numbers
   and may be attributed to the spectral steepening.
We checked if both terms are comparable to each other
   in the observed frequency range (30 MHz $-$ 5 GHz), 
   which corresponds to the wave number range of
   ($10^{-14} \sim  10^{-13}) B_{\mu {\rm G}}^{3/2}$ cm$^{-1}$ (eq.[\ref{k-g}]).
The critical wave numbers $k_{\rm t}$ where
   $(dE/dt)_{\rm accel} = (dE/dt)_{\rm nl}$ are shown in Table \ref{tab3}
   for our best fit models. 
They are certainly within or near the wave number range mentioned above.
The obtained steep wave spectra, $w \sim$ 2.8 and 4.5 for the Coma cluster 
   and $w > 2.0$ in general, could be brought about by this effect.
We suggest that 
   time dependent and self-consistent calculations for non-linear wave-particle systems
   with back reaction of particle acceleration are necessary to investigate
   evolution of the systems.

Next, we make a crude estimate of $P(k)$ in $k<k_{\rm t}$ and
   check the consistency of our consideration mentioned above.
We adopt $P(k)$ of our best fit models for $k>k_{\rm t}$ and
   assume $P(k) \propto k^{-\xi}$ for $k<k_{\rm t}$. 
As a result, the broader band model power spectrum is,
\begin{eqnarray}
   P(k) = \left\{
            \begin{array}{@{\,}ll}
          P_{\rm t} \biggl( \frac{k}{k_{\rm t}} \biggr)^{-\xi}
                  &  \mbox{($k_{\rm min} < k < k_{\rm t}$)} \\
          P_{\rm t} \biggl( \frac{k}{k_{\rm t}} \biggr)^{-w}
                  &  \mbox{($k_{\rm t} < k$)}
            \end{array}
          \right. ,
\end{eqnarray}
   where $P_{\rm t} \equiv b_{\rm t} (B_0^2/8 \pi)(w-1)/k_{\rm t}$ and
   $k_{\rm min}$ is the wave number corresponding to the scale of the cluster,
   $k_{\rm min} \sim 10^{-24}$ cm$^{-1}$.
Then, the total energy density of Alfv\'{e}n waves $M_{\rm tot}$ is given by
\begin{eqnarray}
  M_{\rm tot} =  \frac{w-1}{\xi-1} b_{\rm t}  \frac{B_0^2}{8 \pi}
                 \biggl(\frac{k_{\rm t}}{k_{\rm min}} \biggr)^{\xi-1}
                 \biggl\{
                 1 - \frac{w-\xi}{w-1}
                 \biggl( \frac{k_{\rm t}}{k_{\rm min}} \biggr)^{1-\xi}
                 \biggr\},
\end{eqnarray}
   where we assumed $\xi > 1$.
If we assume
    $M_{\rm tot}$ is
    as large as the energy density of the large-scale-field $B_0^2/(8\pi)$,
    then we obtain $\xi$ with given $k_{\rm t}$ and $b_{\rm t}$.
When $B_0=1\ \mu$G, $k_{\rm t}\sim10^{-14}$ cm$^{-1}$
   and $b_{\rm t} \sim 10^{-6}$,
   we have $\xi \sim 1.6$,
   which is close to
   the index of the energy spectrum, $3/2$,
   formed by the non-linear interaction.
Therefore,
   this result
   is consistent with the previous picture
    of the non-linear interaction for $k < k_{\rm t}$
     and the acceleration reaction for $k > k_{\rm t}$.
When $B_0=0.1\ \mu$G, $k_{\rm t}\sim10^{-15}$ cm$^{-1}$
   and $b_{\rm t} \sim 10^{-3}$,
   we have $\xi \sim 1.2$.

The rate of the energy transfer from $k_{\rm 30MHz}$ to higher wave numbers is estimated as 
   $\varepsilon \sim [P(k_{\rm 30MHz})]^2 k_{\rm 30MHz}^3 / (\rho v_{\rm A})$.
Table \ref{tab3} shows that the values of $\varepsilon$ are less than
   the thermal X-ray cooling rate except for the unrealistic case 
   with $B_0=0.1~\mu$G, $w=4.5$ and $\nu_{\rm c} = 10$MHz.
Therefore, even if the energy of the turbulent Alfv\'{e}n waves goes into the thermal energy of the hot gas
   by the dissipation at higher wave numbers,
   the hot gas could not be heated up.

\subsection{The Origin of the Turbulence}
The energy transfer rate $\varepsilon_0$ at $k_{\rm min} \sim 10^{-24}$ cm$^{-1}$
   is estimated as 
   $\varepsilon_0 \sim P_0^2 k_{\rm min}^3 /\rho v_{\rm A} 
   \sim 10^{-31} B_{\mu {\rm G}}^3$ [erg cm$^{-3}$ s$^{-1}$],
   where $P_0 \sim B_0^2/(8 \pi k_{\rm min})$.
Thus, an energy input rate higher than $ \varepsilon_0$ is necessary to excite and maintain
   the turbulence.
It is certain that motion of galaxies through the intracluster medium
   can excite turbulent eddies.
In this case, however,
   the energy input rate of the steady turbulence is
   \[
    \epsilon_{\rm input} \sim
                         \rho v_0^3 / l_0
                         \sim 10^{-33}
                         \  \mbox{erg} \mbox{ s}^{-1} \mbox{ cm}^{-3},
   \]
   where $l_0 \sim 10$ kpc is a dimension of the galaxy,
   and $v_0 \sim 3.2$ km s$^{-1}$ is the turbulent velocity at $l_0$
   (Goldman \& Rephaeli 1991).
This input rate is insufficient for the required $\varepsilon_0$.
If the energy of the turbulence is input
   by the system of scale $R \sim 1$ Mpc
   with the velocity comparable to the sound velocity
   $v_{\rm s} \sim 1000$ km s$^{-1}$,
   the energy input rate is
   \begin{eqnarray}
     \epsilon_{\rm input} &\sim&
                          \frac{\rho v_{\rm s}^2}
                               {R/v_{\rm s}} \nonumber \\
                          &\sim&
                           10^{-28}
                           \left( \frac{n}{10^{-3} \mbox{cm}^{-3}}\right)
                           \left( \frac{v_{\rm s}}
                                       {1000\mbox{ km s}^{-1}} \right)^3 \nonumber\\
                          &    &
                           \times
                           \left( \frac{R}{1\mbox{ Mpc}} \right)^{-1} 
                           \mbox{ erg} \mbox{ s}^{-1} \mbox{ cm}^{-3}.
   \end{eqnarray}
This input rate
   is sufficient for the radio halo formation.
A merging event between sub-structures can input the kinetic energy
   into the turbulence at this high rate.
Moreover, cluster merger can produce seed CR
electrons through 1st order Fermi acceleration at shocks in
ICM (Takizawa \& Naito 2000). 
Indeed, recent high resolution MHD 
simulations of cluster mergers (Roettiger et al. 1999) show 
that the bulk flow is replaced by turbulent motion in the
later stages of the merger. 

\subsection{The Supply of Cosmic Ray Electrons}
We calculate the total number of the CR electrons $N_{\rm CR}$ from the luminosity
   and find $N_{\rm CR} \sim 10^{60}$ for the Coma radio halo,
   where we adopt the luminosity $L_{\rm t} \sim 10^{41}$ erg s$^{-1}$
   (10 MHz $ \leq \nu \leq $ 10 GHz), the spectral index $\alpha \sim 1.3$,
   and the magnetic field strength $B_0 \sim 1~\mu$G.

We examine how much the radio galaxy can supply the CR electrons.
NGC4874 is one of the two dominant galaxies in the Coma cluster.
The observed flux of the radio source associated with NGC4874 is
   $S_{\rm 408MHz} \sim 4 \times 10^{-24}$ erg cm$^{-2}$ s$^{-1}$ Hz$^{-1}$
   at 408 MHz (Jaffe \& Perola 1974).
From the observed flux  we calculate the number density of the CR electrons
   at the injection energy $\sim 100$ MeV.
For simplicity, we assume that
   the shape of the source is a sphere with a radius $r \sim 10$ kpc,
   and that the radio spectrum follows a single power law with an index $\alpha \sim 0.5$.
We also assume the magnetic field strength is 10 $\mu$G.
We find the number density $n_{\rm CR} \sim 6 \times 10^{-11}$ cm $^{-3}$ at 100 MeV.
The injection rate of the CR electrons leaked from the radio source is given by
   $\dot{n}_{\rm CR} \sim 4 \pi r^2 n_{\rm CR} v_{\rm D}$,
   where $v_{\rm D}$ is the diffusion velocity of the CR electrons.
Taking $v_{\rm D} \sim v_{\rm A} \sim 700$ km s$^{-1}$ gives an upper limit of the injection rate
   $\dot{n}_{\rm CR} \lesssim 4 \times 10^{43}$ s $^{-1}$.
When  $\dot{n}_{\rm CR} \sim 4 \times 10^{43}$ s $^{-1}$,
   the CR electrons in the radio halo are supplied in
   $N_{\rm CR}/\dot{n}_{\rm CR} \sim 7 \times 10^8$ yr.

The merger shock acceleration supplies the CR electrons in the radio halo region.
Takizawa \& Naito (2000) assumed that the total kinetic energy of accelerated electrons
   is 5 \% of the viscous energy, which is nearly equal to the energy of the shock heating,
   and they found that the luminosity of the synchrotron radio emission (10 MHz$-$10 GHz)
   is $\sim 10^{43}$ erg s$^{-1}$.
We find the number of the CR electrons calculated from the luminosity is $5 \times 10^{62}$
   and sufficient for the observed radio halo,
   where we assume the magnetic field strength is $0.1~\mu$G and the radio spectral index is 0.7.
According to Takizawa \& Naito (2000), the radio luminosity decreases after the most contracting epoch.
We suggest that the turbulent Alfv\'{e}n waves reaccelerate the CR electrons injected by the merger shock
   and maintain the radio halo.


\section{SUMMARY}
We have examined
   whether the particle acceleration by the turbulent Alfv\'{e}n waves
   produces the radio halo in the cluster of galaxies.
As a result,
   the energy spectrum of the turbulent Alfv\'{e}n waves
   reproducing the Coma radio spectrum has $w \sim 2.8$.
It is most likely that 
   the turbulence with sufficient energy for the radio halo formation 
   is originated from some merging events between sub-structures.
In our steady model,
   evolution of the turbulence and the magnetic fields
   with the cluster evolution
   is not considered.
By considering effects of the cluster evolution,
   difference between the halo clusters and the non-halo clusters
   would be clarified.
Such difference may also be indicated
   by observation of intracluster hard X-ray and thermal 
X-ray.


\acknowledgments
We thank T. Terasawa, M. Tosa, and M. Hattori for helpful suggestions.
We would also like to thank the anonymous referee for his/her very useful comments.

\clearpage



\clearpage

\clearpage

\begin{deluxetable}{lcccccccc}
\tabletypesize{\scriptsize}
\tablecaption{The calculated spectral index with $w$ and $\nu_{\rm c}$. \label{tab1}}
\tablewidth{0pt}
\tablehead{
\colhead{$\nu_{\rm c}$} & \multicolumn{8}{c}{$w$}  \\
\cline{2-9}\\
\colhead{(MHz)}&\colhead{2.0}&\colhead{2.6}&\colhead{2.8}&\colhead{3.2}&
\colhead{3.5}&\colhead{4.0}&\colhead{5.0}&\colhead{6.0}
}
\startdata
 $10^4$          &   P   &   P   &   F   &  S    & S     &  S   & S    &  S   \\
 $10^3$          &   P   &   F   &   F   & 2.07  & S     &  S   & S    &  S   \\
 $10^2$          & 1.02  & 0.97  & 1.03  & 1.35  & 2.61  &  S   & S    &  S   \\
 $10$            &  S    & 1.90  & 1.44  & 0.96  & 1.16  &  1.79& S    &  S   \\
 $1$             &  S    &  S    & 2.03  &   F   & 0.88  &  1.08& 1.54 &  S   \\
 $0.1$           &  S    &  S    &  S    &   F   &   F   &  1.02& 1.50 &  2.00\\
 $0.01$          &  S    &  S    &  S    &   F   &   F   &  1.00& 1.50 &  2.00  
\enddata 
\tablecomments{Values in columns are $\alpha$ between 30 MHz and 5 GHz.
               Spectra of $\alpha>2$ and spectra of $\alpha < 1$
               are represented by the character `S' and `F', respectively.
               Spectra which has a peak in the observed frequency range
               is represented by `P'.}
\end{deluxetable}

\clearpage

\begin{deluxetable}{lccccccccc}
\tabletypesize{\scriptsize}
  \tablecaption{The calculated values of $J$ with $w$ and $\nu_{\rm c}$. \label{tab2}}
\tablewidth{0pt}
\tablehead{
 \colhead{$\nu_{\rm c}$} & \multicolumn{9}{c}{$w$} \\
 \cline{2-10}\\
 \colhead{(MHz)}&\colhead{2.0}&\colhead{2.6}&\colhead{2.8}&\colhead{3.2}&
 \colhead{3.5}&\colhead{4.0}&\colhead{4.5}&\colhead{5.0}&\colhead{6.0}\\
}
\startdata
 $10^4$          &\nodata&\nodata&\nodata&\nodata&\nodata&\nodata&\nodata&\nodata&\nodata\\
 $10^3$          & 0.71 & 0.49 & 0.30  & 0.33  &\nodata&\nodata&\nodata&\nodata&\nodata\\
 $10^2$          & 0.14 & 0.11 & 0.075 & 0.053 & 0.91   &\nodata&\nodata&\nodata&\nodata\\
 $10$            & 1.4  & 0.14 & 0.018 & 0.11  & 0.070 & 0.17 &\nodata &\nodata&\nodata\\
 $1$             &\nodata& 1.6  & 0.20  & 0.18  & 0.13  & 0.063& 0.027 &0.036 &\nodata \\
 $0.1$           &\nodata&\nodata& 1.0   & 0.22  & 0.16  & 0.077& 0.028 &0.030 & 0.20 \\
 $0.01$          &\nodata&\nodata&\nodata& 0.24  & 0.18  & 0.082& 0.028 &0.029 & 0.20     
\enddata
\end{deluxetable}

\clearpage

\begin{deluxetable}{lcccccccc}
\tabletypesize{\scriptsize}
\tablecaption{Model parameters reproducing the observed radio spectrum. \label{tab3}}
\tablewidth{0pt}
\tablehead{
 \colhead{} &
 \colhead{} &
 \colhead{} &
 \multicolumn{2}{c}{$q$ \tablenotemark{c}} &
 \colhead{} &
 \colhead{} & 
 \colhead{} \\
 \colhead{} &
 \colhead{} &
 \colhead{$b B_0^2/(8 \pi)$ \tablenotemark{b}} &
 \multicolumn{2}{c}{(cm$^{-3}$ s$^{-1}$)} &
 \colhead{$k_{\rm t}$ \tablenotemark{d}} &
 \colhead{} & 
 \colhead{$\varepsilon$ \tablenotemark{f}} \\
 \colhead{($w$, $\nu_{\rm c}${\tiny [MHz]})} &
 \colhead{$b$ \tablenotemark{a}} &
 \colhead{(erg cm$^{-3}$)} &
 \colhead{{\tiny $E_{\rm i}$=100 MeV}} &
 \colhead{{\tiny $E_{\rm i}$=1 GeV}} &
 \colhead{(cm$^{-1}$)} &
 \colhead{$b_{\rm t}$ \tablenotemark{e}} & 
 \colhead{(erg cm$^{-3}$ s$^{-1}$)} \\
}
\startdata
$B_0=1~\mu $G &&&&&&&&\\
(2.8, 10) & $1.7 \times 10^{-5}$ & $6.8 \times 10^{-19}$ & $3.8 \times 10^{-29}$ & $2.3 \times 10^{-30}$ &
           $5.7 \times 10^{-14}$ & $6.1 \times 10^{-7}$  & $2.2 \times 10^{-33}$ \\  
(4.5, 1.0)& $1.5 \times 10^{-2}$ & $6.0 \times 10^{-16}$ & $2.9 \times 10^{-9}$ & $4.1 \times 10^{-29}$ &
           $1.5 \times 10^{-13}$ & $8.1 \times 10^{-7}$  & $1.8 \times 10^{-30}$ \\ 
$B_0=0.1~\mu $G &&&&&&&&\\
(2.8,10)  & $1.6 \times 10^{-1}$ & $6.2 \times 10^{-17}$ & $1.5 \times 10^{-30}$ & $1.4 \times 10^{-30}$ &
           $4.6 \times 10^{-15}$ & $1.0 \times 10^{-3}$  & $5.7 \times 10^{-30}$ \\
(4.5,1.0) & 140 & $5.5 \times 10^{-14}$ & $1.1 \times 10^{70}$ & $2.5 \times 10^{-26}$ &
           $8.9 \times 10^{-15}$ & $8.3 \times 10^{-4}$  & $5.0 \times 10^{-27}$\\
\enddata

\tablenotetext{a}{The value $b$ is the fractional energy density of Alfv\'{e}n waves
                  above $k_{\rm 5GHz} = 9 \times 10^{-15} B_{\mu{\rm G}}^{3/2}$ cm$^{-1}$.}
\tablenotetext{b}{The energy density of Alfv\'{e}n waves above $k_{\rm 5GHz}$.}   
\tablenotetext{c}{The values of the injection rate $q$($= 4 \pi p_{\rm i}^2 q_{\rm i}$) 
                  for the steady state solution to reproduce the radio intensity.}
\tablenotetext{d}{The value $k_{\rm t}$ is the wave number at which
                  the energy loss rate due to the acceleration overcomes
                  the energy transfer rate via the nonlinear interaction.}
\tablenotetext{e}{The value $b_{\rm t}$
                  is the fractional energy density above $k_{\rm t}$.}
\tablenotetext{f}{The value $\varepsilon$ is the energy transfer rate
                  at $k_{\rm 30MHz}$.}
\end{deluxetable}
\end{document}